\documentclass[aps,prd,amsmath,amssymb,superscriptaddress,nofootinbib,10pt]{revtex4}
\usepackage{hyperref}
\usepackage{amssymb}
\usepackage{bm}
\usepackage{latexsym}
\usepackage{amsfonts,amssymb}
\usepackage{graphicx,epsfig}
\usepackage{psfrag}
\usepackage{amsmath,amssymb}
\usepackage{mathrsfs}

\usepackage{bm}
\usepackage{latexsym}
\usepackage{mathrsfs}

\newcommand{\addlabel}[1]{%
\refstepcounter{equation}%
\addlabel{#1}%
\let\]\endequation }

\def\MPlh{M_{_{\rm P}}}

\begin{document}

\title{Inflation with $f(R,\phi)$ in Jordan frame}

\author{Jose Mathew}\email{jose@iisertvm.ac.in}
\affiliation{School of Physics, Indian Institute of Science Education
  and Research Thiruvananthapuram (IISER-TVM), Trivandrum 695016,
  India}
\author{Joseph P Johnson} \email{josephpj@iitb.ac.in}
\author{S. Shankaranarayanan} \email{shanki@phy.iitb.ac.in} \affiliation{Department of Physics, Indian Institute of Technology Bombay, Mumbai 400076, India}


\begin{abstract}
We consider an $f(R)$ action that is non-minimally coupled to a
 massive scalar field. 
The model closely resembles scalar-tensor
 theory and by conformal transformation can be transformed to
 Einstein frame. To avoid the ambiguity of the frame dependence, we
 obtain an exact analytical solution in Jordan frame and show
 that
 the model leads to a period of accelerated expansion with an
 exit. Further, we compute the scalar and tensor power spectrum for
 the model and compare them with observations.
\end{abstract}

\maketitle
\section{Introduction}
Cosmological Inflation ~\cite{book:15483,book:15485,book:15486,book:75690,mukhanov1992theory} was originally introduced in the early 
1980s to solve the cosmological puzzles like the horizon problem, flatness 
problem of the FRW model. It is now considered as the best paradigm for describing the early stages of the universe as it can explain the origin of structures in the universe and anisotropies in the Cosmic Microwave Background (CMB). The current PLANCK measurements indicate that the temperature fluctuations of CMB are 
nearly scale invariant~\cite{ade2014planck}. Hence, the success of an inflationary 
model not only rests on providing a minimum of 50 e-foldings of inflation required 
to solve the cosmological puzzles, but a highly demanding requirement that the 
predicted power-spectrum being close to scale invariance~\cite{lyth1999particle,lyth2008particle,mazumdar2010particle,yamaguchi2011supergravity}  .

While inflation is the successful paradigm for the early universe, we do not yet have a fundamental understanding of the mechanism that drives inflation. Inflationary models based on canonical scalar field built within General Relativity, need to satisfy slow-roll conditions~\cite{lidsey1997reconstructing,bassett2006inflation}. In other words, the observations require the scalar field potentials to be nearly flat~\cite{ade2014planck}.
For the quantum fluctuations that exit the horizon during inflation to be nearly scale-invariant, the energy scale of the universe must remain almost a constant which require the canonical scalar field potential to be almost flat -- almost like cosmological constant. While these flat potentials are phenomenologically successful, the scalar field sector of action deviates much from the standard model of particle physics. For instance, the
renormalizability puts a constraint on the scalar field potential to be quartic i.e. $V(\phi) = m^2 \phi^2 + \lambda \phi^4$, where $\lambda$ is the coupling parameter and m is the mass. The inflationary models, based on canonical scalar field, require potentials of the form $V(\phi) = \sum_{n = 0}^{N}
c_{2n} \phi^{2n}$ where $c_{2n}$'s are real numbers and $N > 2$. However, in the standard model of particle physics and its minimal extensions, there is no candidate for inflaton with such flat potentials that could sustain inflation~\cite{lyth1999particle,lyth2008particle,mazumdar2010particle,yamaguchi2011supergravity}.  In this work, we construct an inflationary model where the inflaton is a massive scalar field which is theoretically well placed in the standard model of particle physics,  however, in achieving this, we go beyond General Relativity.  

The Einstein-Hilbert action can be treated as a limiting case of a more general action containing higher
order invariants, and $f(R)$ gravity assumes such an action to be a function of Ricci scalar\cite{sotiriou2010f,de2010f,clifton2011modified,nojiri2011unified}. General Relativity is not renormalizable and, therefore, can not be quantized
conventionally. However, the Einstein-Hilbert action supplemented with higher order
curvature terms is renormalizable \cite{sotiriou20096+,benedetti2009asymptotic} which makes $f(R)$ gravity an interesting alternative
to General Relativity. As usual, this comes with a prize, unlike General Relativity the
resulting field equations of $f(R)$ gravity are not second order, but fourth order, which makes it non-trivial. $f(R)$ theories do not suffer from Ostr\"ogradsky instability~\cite{2007-Woodard-Proc}.

 In this work, we consider $f(R,\phi)$ \cite{RADOR2007228,de2016spotting} action, where $f(R)$ is
 non-minimally
coupled to a massive scalar field, closely resembling
 scalar-tensor theories of gravity \cite{fujii2003scalar}. As
 mentioned above, the potential is fixed to be $m^2 \phi^2$ and we
 choose the coupling function such that the modification to the
 Einstein-Hilbert action is dominant only at the initial phase of
 inflation. It is possible to perform a conformal transformation to
 the $f(R,\phi)$ action and transform it to Einstein-Hilbert action
 with an extra scalar field~\cite{mukhanov1992theory}. 

Scalar-tensor theories of gravity have a long-standing controversy about which frame (Einstein or Jordan) is the physical one \cite{1994-Magnano.Sokolowski-PRD,2004-Flanagan-PRL,2006-Multamaki.Vilja-PRD}. To avoid these controversies and the ambiguity of the frame dependence, in this work, we consider the action without
performing any identification or transformation with any other theory, frame or variables~\cite{2006-Multamaki.Vilja-PRD}. More specifically, we obtain an exact analytical solution in Jordan frame and show that the model leads to a period of accelerated expansion with an exit. We obtain the power spectrum for this model and show that it is nearly scale invariant. 

The paper is organised as follows: In the next section, we discuss the model 
and obtain the exact analytical solution. To physically understand the dynamical equations, we obtain the exact de Sitter solution. In Sec. \ref{sec:specialcase_themodel}, we discuss our model and show
that the constant $H$ solution is a saddle point solution. We use this feature to highlight that the model has a graceful exit. In Sec. \ref{sec:Pspectrum}, we compute the power-spectrum of our model. Further, we discuss the key results and possible implications of our model in Sec. \ref{sec:Conclusions}.

In this work, we consider $(-, +, + ,+)$ metric signature. We use lower Latin alphabets for the 4-dimensional space-time and lower Greek alphabets for the 3-dimensional space. We use natural units $c=\hbar=1$, $\kappa=1/\MPlh^2$, and $\MPlh$ is the reduced Planck mass. We denote dot as derivative with
respect to cosmic time $t$ and $H(t) \equiv \dot{a}(t)/a(t)$. Various physical quantities with the overline refers to the values evaluated for the homogeneous and isotropic FRW background.

\section{Model and exact background solution}
\label{sec:exact_solution}
We consider the following action:
 \begin{equation}\label{eq:action}
S =  \int  d^{4}x \, \sqrt{-g} \left[\frac{1}{2} f(R,\phi)-
 \frac{\omega}{2}   g^{a b} \nabla_{a} \phi \nabla_{b} \phi
 -V(\phi)\right] ,
 \end{equation}
where $\phi$ is the scalar field, $V(\phi)$ is the scalar field potential and, we assume 
\begin{equation}
f(R,\phi)=h(\phi)\left(R+\alpha R^2\right)
\end{equation}
$h(\phi)$ is non-minimal coupling function. It is important to note that $\omega = 1$ corresponds to canonical scalar field while $\omega = - 1$ corresponds to ghost~\cite{2004-Arkani-Hamed.etal-JCAP}. Varying action (\ref{eq:action}) w.r.t. the 
field $\phi$ and the metric leads to the following equations of motion \cite{hwang2000conserved}:
\begin{eqnarray}
\label{eq:PhiEOM}
& & \Box \phi +  \frac{1}{2
 \omega}\left(\omega_{,\phi}\phi^{;a}\phi_{,a}+f_{,\phi} -2V_{,\phi}\right) = 0 \\
 && F \, G^{p}_{q}=\omega\left(\phi^{;p}\phi_{;q}-\frac{1}{2} \delta^{p}_{q} \phi^{;c}\phi_{;c}\right) -\frac{1}{2}\delta^{p}_{q}\left(RF-f+2V\right)+F^{;p}_{\;\;q} - \delta^{p}_{q}\Box{F}
 \label{eq:EinsEOM}
\end{eqnarray}
where $F  = \partial f(\phi,R)/\partial R$. The stress-tensor of the scalar field and 
modified gravity are given by 
\begin{subequations}
\label{eq:stresstensor}
\begin{eqnarray}
T^{\phi}_{pq} &=& \omega \left( \partial_{\mu} \phi  \partial_{\nu} \phi 
- \frac{1}{2} g_{pq} \partial_{c} \phi  \partial^{c} \phi  \right)
- g_{pq} V(\phi) \\
T^{MG}_{pq} &=& \frac{g_{pq}}{2} \left(f - F \, R \right)
+ \nabla_{p} \nabla_{p} F - g_{pq} \Box F
\end{eqnarray}
\end{subequations}
 In the rest of this section, we obtain exact solution for the FRW
 background in the Jordan frame (without performing conformal
 transformation). 

\subsection{Exact background solution in Jordan frame}

As mentioned earlier, to avoid ambiguity, we obtain the exact solution for the 
spatially flat Friedmann-Robertson-Walker (FRW) background:
\begin{equation}
ds^2=-dt^2 + a^2(t) \left( dx^2 + dy^2 + dz^2 \right)
\end{equation}
in the Jordan frame. Note that $a(t)$ is the scale factor. From 
Eqs. (\ref{eq:PhiEOM},\ref{eq:EinsEOM}), equation of motion for 
$\phi(t)$ and the Hubble parameter, $H(t)$, are:
\begin{subequations}
\label{eq:threeequations}
\begin{eqnarray}
\label{eom}
6\dot{h}H^2+72\dot{h}H^4\alpha+72\dot{h}H^2\alpha\dot{H} + 3\dot{h}\dot{H}+18\dot{h}\dot{H}^2\alpha-\dot{V}-\omega \, \dot{\phi} \ddot{\phi}-3\omega H \dot{\phi}^2 &=& 0 \\ 
\label{energyconstraint}
-\frac{1}{2}\omega \dot{\phi}^2 + {3hH^2}+108\alpha hH^2\dot{H}-18h\dot{H}^2 \alpha-V+3H\dot{h} + 72 \dot{h}\alpha H^3 +36 H\dot{h}\alpha\dot{H} + 36 H h\alpha \ddot{H}&=& 0 \\
\label{evolutionequation}
2h\dot{H}+108 \alpha hH^2\dot{H} + 48 \dot{h}H^3 \alpha+54h\dot{H}^2\alpha +3hH^2 +\ddot{h}+\frac{1}{2}\omega \dot{\phi}^2 120 H \dot{h}\dot{H}\alpha + 72h\ddot{H}\alpha + 2H\dot{h}& &\nonumber\\
-V+24\alpha H^2\ddot{h} +12\alpha\dot{H}\ddot{h} +24\alpha\ddot{H}\dot{h}+12h\alpha\dddot{H}&=& 0
\end{eqnarray}
\end{subequations}
Rewriting Eqs.~(\ref{eom}, \ref{energyconstraint}), we get
\begin{eqnarray}
-2h\dot{H}-72h\dot{H}^2\alpha-\omega \dot{\phi}^2 -\ddot{h}-84H\dot{h}\dot{H}\alpha-36Hh\ddot{H}\alpha + 24 \alpha \dot{h} H^3 +H\dot{h}-24\ddot{h}H^2\alpha - 12\alpha \ddot{h}\dot{H} -24\alpha\dot{h}\ddot{H}-12\alpha h  \dddot{H}=0
\label{eq:keyeqn}
\end{eqnarray}
As it is evident, the equations of motion are higher order. To obtain the exact solution to the above equations  (\ref{eq:threeequations},\ref{eq:keyeqn}), we use the  following ansatz:
\begin{equation}
\label{eq:dsansatz}
a(t) =a_0 e^{H_D t}~~\mbox{and}~~ \phi=\phi_0 e^{-p \, H_Dt} 
\end{equation}
where $a_0$, $H_D > 0$ and $n$ are constants. $\phi_0$ corresponds to the value 
of the scalar field at the start of inflation which is set to $1$.  
Substituting the above ansatz (\ref{eq:dsansatz}) in
Eq.~(\ref{eq:keyeqn}) and solving for $h(\phi)$ \cite{mathew2016low}, we get the following {\sl exact relations}:
\begin{subequations}
\label{eq:dssol}
\begin{eqnarray}
  \label{eq:PLgensol1}
V\left(\phi\right) &=& \;\lambda_0
\!\!\! \; \; + \; m^2 \; {\phi}^{2}
\!\!\! \; \; \; + \; {\lambda}_p \; {\phi}^{-p}   \\ 
\label{eq:PLgensol2}
h\left(\phi\right) &=& \; \mu_0  
\; + \; {\mu}_2 \; {\phi}^{2}
\; + \; {\mu}_p \; {\phi}^{-p}
\end{eqnarray}
\end{subequations}
where 
\begin{eqnarray}
\mu_0&=& \frac{1}{3\, H_D^2} \lambda_0 \\
\mu_2&=&- {\frac {\omega\,
 p}{ \left(1 + 24\,\alpha\,H_D^{2} \right) \left( 2+ 4\,p \right) }}\\
m^2&=&   \left(3 + p (2 p - 5) (1 + 24 \,\alpha\,H_D^{2}) \right) \, H_D^2 \, \mu_2\\
\mu_p & =&   \frac{1}{6\, \,H_D^{2} \left( 12\,\alpha\,H_D^{2}+1 \right)} \lambda_p \, ,  
\end{eqnarray}
$\lambda_0, \lambda_p$ are arbitrary (integration) constants. It is important to note that this is an exact analytical expression for the background evolution. We like to stress the following points: 
First, as mentioned earlier, the solution is obtained without 
performing any conformal transformation. To our knowledge, there has been no exact inflationary solution in the Jordan frame.  
Second, for the exact de Sitter, the scalar field decays with time. This has 
been contrasted with general relativity where the scalar field is a constant.
Third, the coupling function is strongly related to the potential. From the above expressions, it is clear that $\mu_0$ depends on $\lambda_0$, similarly, $\mu_p$ depends on $\lambda_p$ and $m$ is related to $\mu_2$. If $\lambda_0$, $\lambda_p$ vanish, then automatically, corresponding $\mu_0, \mu_p$ also vanish. This is important because, while the scalar field potential can be obtained from the standard model of particle physics, the non-minimal coupling term is  completely independent. Since each of the terms in the non-minimal coupling term is related to the scalar field potential, for our exact model, the non-minimal coupling term is completely fixed by the standard model of particle physics. 
Lastly, the anatz (\ref{eq:dsansatz}) is for exact de Sitter leading to an accelerated expansion. Since the field decays exponentially (\ref{eq:dsansatz}), the non-minimal coupling function and the scalar field potential are dominated by $\phi^{-p}$. This implies that once the inflation sets in, there is no way to stop the inflation if $\lambda_p \neq 0$. As mentioned in the introduction, our focus is to have less deviation from the standard model of particle physics, hence, we set $\lambda_0 = \lambda_p = 0$. Although $\lambda_0 = 0$ is not a requirement, this assumption simplifies our analysis as we will show in the next subsection.

\subsection{Special case: $\lambda_0\;=\;\lambda_p \; = \;0$}
\label{sec:specialcase_themodel}
As mentioned above, let us set $\lambda_0\;=\;\lambda_p \; = \;0$ and $\omega=1$. We have:
\begin{eqnarray}
\label{eq:dssplsol}
 V\left(\phi\right)=  m^2 {\phi}^{2} &; & 
 h\left(\phi\right)= \mu_2 \phi^2 
\end{eqnarray}
where 
\begin{equation}
\label{eq:dsspcoefficients}
\mu_2 = - {\frac {p}{ \left(1 + 24\,\alpha\,H_D^{2}\right) \left( 2+
 4\, p \right) }} \quad ; \quad m^2 =  \left(3 + p (2 p - 5) (1 + 24 \,\alpha\,H_D^{2}) \right) 
H_D^2 \, \mu_2 \\
\end{equation}
This is one of the main results of this work regarding which we would like to stress the following points: 
First, while $\lambda_p$ is set to zero, $p$ is an arbitrary number and can take any positive value. 
Second, $\mu_2$ is directly related to $m^2$. Since, $m^2$ is positive definite, this leads to the condition that ${24 \alpha H_D^2}+1<0$ or $\alpha < -1/(24 H_D^2)$. Since the Hubble parameter during inflation is large, this implies that $\alpha$ is small negative value. 
Third, to understand the exact analytical solution, from the stress-tensor (\ref{eq:stresstensor}), let us calculate $\rho + 3 P$ 
 \begin{equation}
\rho + 3 P \equiv - T^0_0 + T^\alpha_\alpha  =   
\left[p^2 - \frac{m^2}{H_D^2} 
 + \mu_2 \, \left[ 180 \, \alpha H_D^2 
 + 3 \, p \, (2 p - 1)  (1 + 24 \alpha H_D^2) \right] \right] 
2 \phi_0^2 H_D^2 e^{-2 p H_D t} 
 \end{equation}
The first two terms correspond to the canonical scalar field while
 the last term correspond to the modifications to the gravity. For $p
 > 1/2$, the third term is always negative, while, for $p < 1/2$ the
 third term can be positive, negative or zero. Large value of $p (\gg
 1)$ does not lead to inflation as $\rho + 3 P = 0$.  However, $p \ll
 1$ leads to $\rho + 3 P < 0$. In the analysis we take $p < 1$. It s
 important to 
note that $p \ll 1$ corresponds to constant scalar
 field.
Lastly, the solution we have obtained is exact de Sitter which does not have an exit. 
The above analysis provides the possibility that if $p$ changes from small value to large value this will lead to an inflation with exit. Other way of looking at this is to change the initial value of $\dot{\phi}$. Using the fact that, $\dot{\phi} \propto p$, it is possible to check what kind of inflationary solutions occur if $\dot{\phi}_{t = 0} \neq \dot{\phi}_{t= 0}^{dS}$. We will study this in the next subsection.

\subsection{Saddle point and exit from inflation}
\label{sec:Exitfrominflation}
In this section we show --- both analytically and numerically --- that the exact de Sitter solution obtained is a saddle point. Specifically, we show that the 
deviation of the initial values from the de Sitter value leads to either a smooth exit from the inflationary phase or to super inflation. Hence there exist a range of initial values for which we have a viable inflationary solution. Models of inflation based on the saddle point solutions have been considered in past, see Refs.~\cite{kanti2015gauss,Rinaldi:2015uvu,Tambalo:2016eqr}.

\subsubsection{Analytical}

\begin{figure}[!htb]
\centering 
\includegraphics[width=0.9 \linewidth]{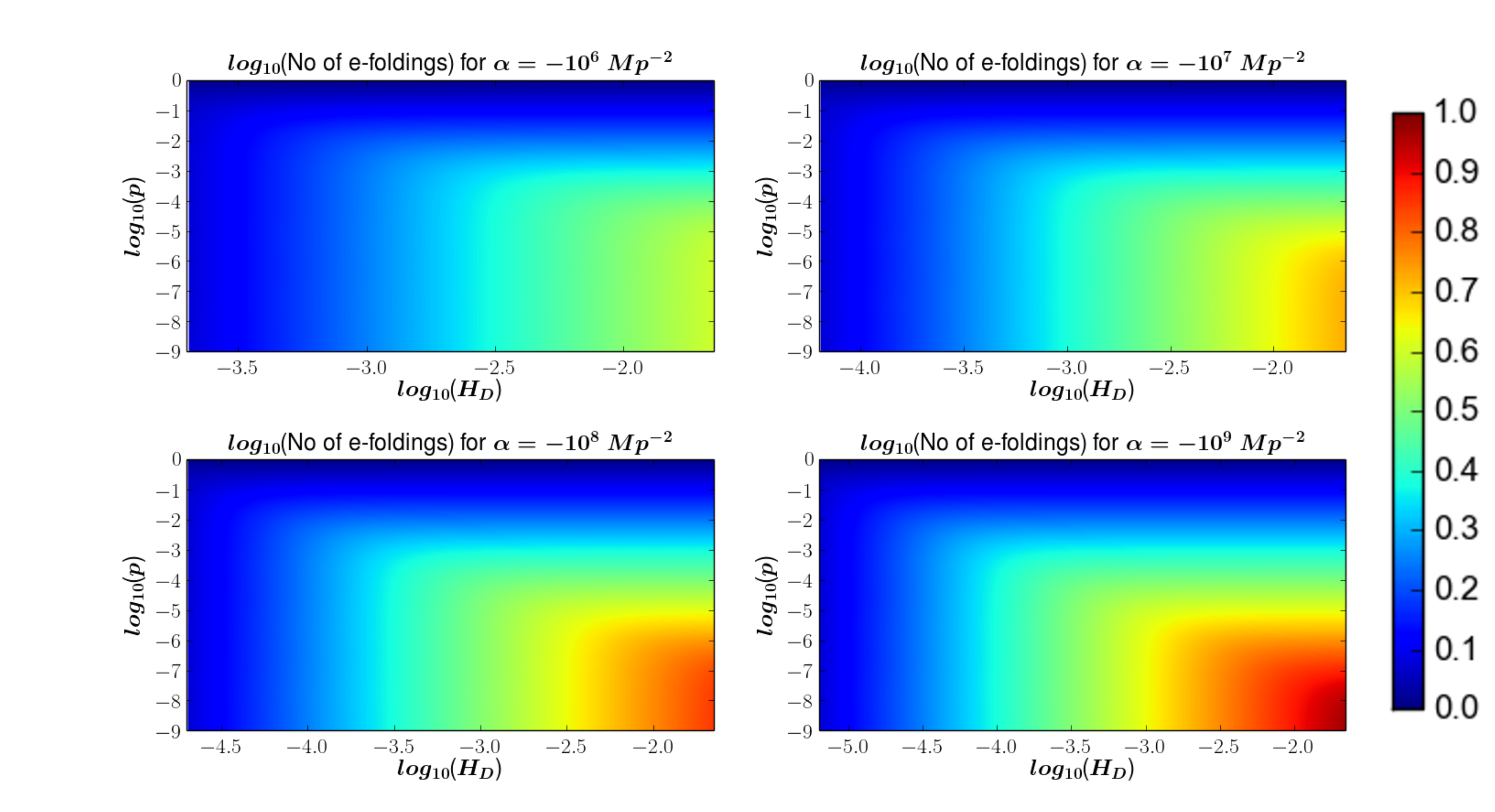}
\caption{Contour plot showing the dependence of the number of
 e-folding on $H_D$, $p$ and $\alpha$. } 
\label{fig:Nonqn}
\end{figure}

Analytically, we look at the trajectories close to the de Sitter solution and evolve them   in time. Rewriting the equations (\ref{eq:threeequations}, \ref{eq:keyeqn}) 
in terms of the variable $\Delta = \dot{\phi}/{\phi}$:
\begin{equation}
\begin{aligned}
\dot{\Delta} &= {144 \alpha \mu_2 H^4} + {144  \alpha
 \mu_2 \dot{H}H^2} + {36 \alpha \mu_2
 \dot{H}^2}-3H\Delta - \Delta^2 +{12 \mu_2
 H^2}+{6\mu_2 \dot{H}} -2m^2 \\
\ddot{H} &= -4H^2 \Delta -3H\dot{H}-2\Delta \dot{H} + \frac{1}{72} \frac{
 \Delta^2}{\alpha \mu_2 H} +\frac{1}{2} \frac{\dot{H}^2}{H} + \frac{1}{36} \frac{m^4}{\alpha \mu_2 H} -\frac{1}{12}\frac{H}{\alpha} - \frac{1}{6}\frac{\Delta}{\alpha}
\end{aligned}
\end{equation}
The fact that we can write the field equations only in terms of $\Delta$ implies that the dynamics (evolution of Hubble parameter, Number of e-foldings etc.) depends only on $\Delta$ and does not depend  on $\phi$ or $\dot{\phi}$ independently. Let us define vector $\bm{v}$ as:
\[
\bm{v} =   \begin{pmatrix} H \\  \dot{H} \\ \Delta \end{pmatrix} 
\]
It is important to note that the de Sitter solution ($H = H_D$) is an equilibrium point ($\{\dot{\bm{v}}\}_{eq}=0$) and 
\[
\{{\bm v} \}_{eq} =   \begin{pmatrix} H_D \\  0 \\ -pH_D \end{pmatrix} \, .
\]
The background equations for $\dot{\bm v}=f({\bm v})$  can be written as 

\[
\dot{\bm v} = 
\begin{pmatrix}\dot{H} \\  \ddot{H} \\ \dot{\Delta} \end{pmatrix}= \begin{Bmatrix} \dot{H} \\ -4H^2 \Delta -3H\dot{H}-2\Delta \dot{H} + \frac{1}{72} \frac{
 \Delta^2}{\alpha \mu_2 H} +\frac{1}{2} \frac{\dot{H}^2}{H} + \frac{1}{36} \frac{m^4}{\alpha \mu_2 H} -\frac{1}{12}\frac{H}{\alpha} - \frac{1}{6}\frac{\Delta}{\alpha}
 \\ {144 \alpha \mu_2 H^4} + {144  \alpha
  \mu_2 \dot{H}H^2} + {36 \alpha \mu_2
  \dot{H}^2}-3H\Delta - \Delta^2 +{12 \mu_2
  H^2}+{6\mu_2 \dot{H}} -2m^2
\end{Bmatrix}\
\]
As mentioned above, we perturb ${\bm v} = {\bm v}_{eq}+ \delta {\bm
 v}$ and obtain the equation for $\delta {\bm v}$ by Taylor expanding $f({\bm v})$ about the equilibrium point. Hence, we have:
\begin{equation}
\dot{\delta v_i}=\{\partial_{j}{f_i}\}_{eq}\delta v_j=J_{i j}\delta v_j \, . 
\end{equation}
See Appendix (\ref{app:analytical}) for details.  From the above analysis, it is clear that the de Sitter solution is a saddle point. Hence, for a range of initial conditions we have an inflationary phase eventually leading to exit. For the largest positive eigen value ($\lambda$), the number of e-foldings is given by: 
\[
N \approx \frac{H_D}{\lambda} ln\left(\frac{H_D^2}{\lambda(H_D-H_i)}\right).
\]
Fig.~(\ref{fig:Nonqn}) contains the contour plot for the  parameters $p$ and $H_D$ for different e-foldings by keeping $(H_D-H_i)/H_D$ constant.
\subsubsection{Numerical}

\begin{figure}[!hbt]
    \centering
    \begin{minipage}{0.5\textwidth}
        \centering
        \includegraphics[width=1\textwidth]{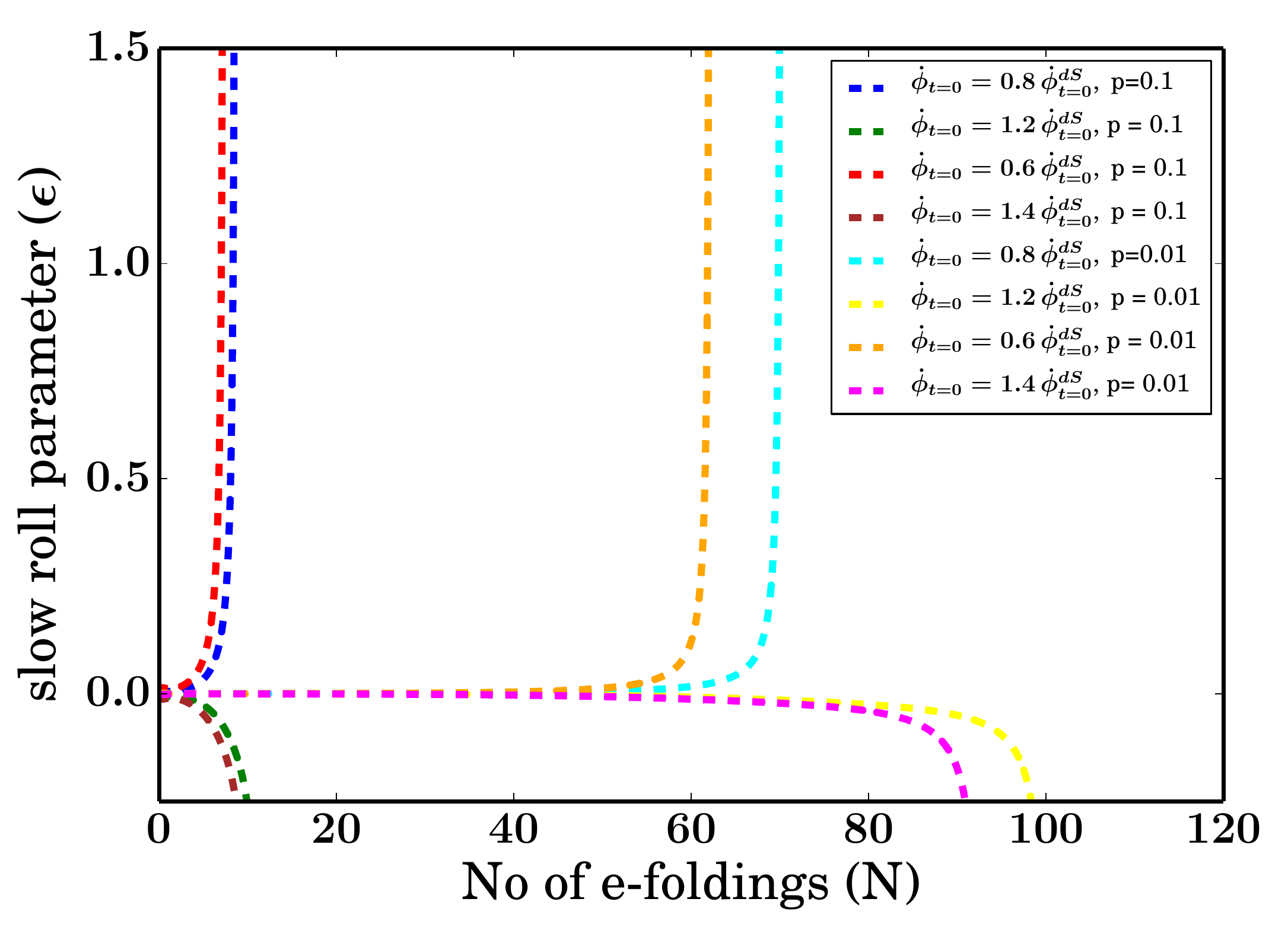} 
       
    \end{minipage}\hfill
    \begin{minipage}{0.5\textwidth}
        \centering
        \includegraphics[width=0.9\textwidth]{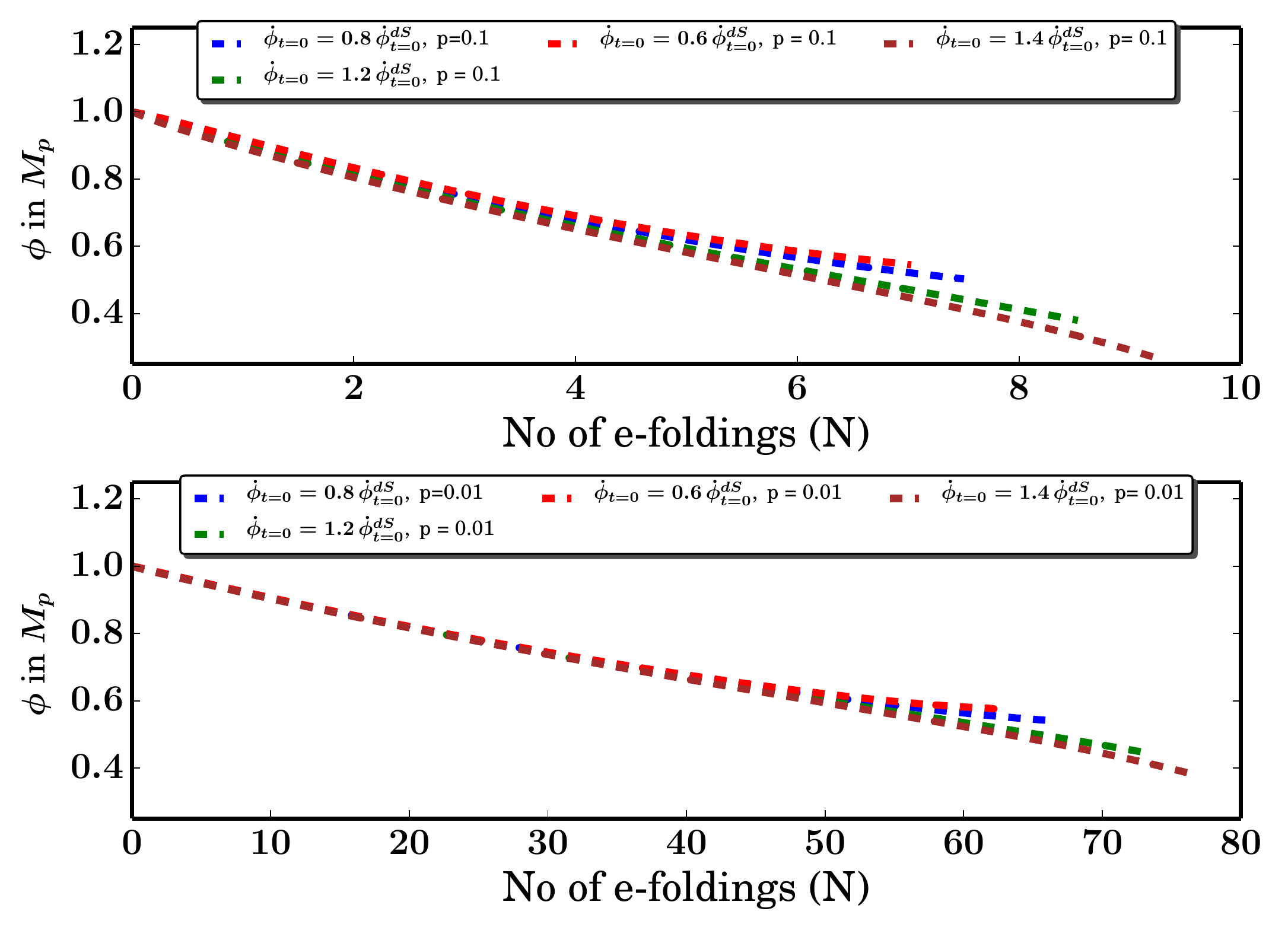} 
    \end{minipage}
    \caption{ Slow roll parameter ${\epsilon}$ and scalar field $\phi$ vs
     number of e-foldings (i) slow-roll parameter, $\epsilon$ and (ii)
     scalar field $\phi$, for different initial values of
     $\frac{\dot{\phi}}{\phi}$.}
    \label{fig:epsilon-phivsN}
\end{figure}

We also studied the evolution of background equations (\ref{eq:threeequations})  numerically for a time step of $10^{-4}\MPlh^{-1}$ and for a precision of the field $\phi/\phi_0$, (in dimensionless units)  $10^{-16}$. Fig. (\ref{fig:epsilon-phivsN}) contains the plot of slow-roll parameter $\epsilon = -\frac{\dot{H}}{H^2}$ and 
the scalar field $\phi$ as a function of the number of e-foldings for different 
initial values of $\dot{\phi}$.  We have taken the values for the parameters to be $\mu_2=10^{-4} \;\mbox{,}\;\;\alpha=-10^{8}M_p^{-2}$ and two different values for $p$, $p=0.1$ and $p=0.01$, that corresponds to $m=6 \times 10^{-5} M_p$, $H_D=4\times 10^{-4} M_p$ and  $m=3.36 \times 10^{-6} M_p$, $H_D=10^{-4}M_p$ respectively.

%
%

From the plots we infer that the number of e-foldings is larger for larger values of $H_D$ and for smaller values of $p$. These are consistent with the analytical 
results as small value for $p$ suggests that the Universe remain in the inflationary phase for a long period.

\section{Scalar and Tensor power Spectra for the model}

\label{sec:Pspectrum}
In this section, we compute the scalar and tensor power-spectrum for our model discussed in the previous section. For easy comparison, we use the same notation  as in Ref.~\cite{hwang1996cosmological}. It is important to note that the analysis used in Ref.~\cite{hwang1996cosmological} is not applicable for our model and hence, the results obtained in Ref.~\cite{2015-Myrzakulov.etal-EPJC} have to be interpreted cautiously. 

\subsection{Perturbations}
The linear order perturbations about the FRW background is given by \cite{hwang1996cosmological}
\begin{equation}
ds^2= -(1+2 \theta)dt^2-a(\beta_{,\alpha}+B_{\alpha}) dtdx^{\alpha} + a^2[g^{(3)}_{\alpha \beta}(1-2 \psi) +2 \gamma_{,\alpha|\beta} +
 2C_{\alpha|\beta} +2C_{\alpha \beta}]
\end{equation}
where $\theta(x,t)$, $\beta(x,t)$, $\psi(x,t)$ and $\gamma(x,t)$ characterize the scalar-type perturbations, $B_\alpha(x,t)$ and $C_\alpha(x,t)$ are trace-free ($B^{\alpha}_{|\alpha} = C^{\alpha}_{|\alpha} = 0$) vector perturbations, and $C_{\alpha \beta}(x,t)$ is transverse, trace-free $(C^{\beta}_{\alpha|\beta}=0=C^{\alpha}_\alpha)$  tensor perturbation. The scalar field is decomposed as $\phi(x,t)=\bar{\phi}(t)+\delta \phi(x,t)$. 

The perturbed scalar equations in Newtonian gauge in  the Fourier space are given by~\cite{hwang1996cosmological,hwang2000conserved}
\begin{subequations}
\begin{eqnarray}
& &-F \psi +F \theta + \delta F = 0\label{eq:pertinotj}\\
& & -2F\dot{\psi} -2FH\theta -\dot{F}\theta +\dot{\phi}\delta \phi
 +\dot{\delta F}-H\delta F = 0 \label{eq:pert0i}\\
& & 6FH\dot{\psi}+6FH^2\theta + 2F\frac{k^2}{a^2}\psi -\dot{\phi}^2
 \theta + 3 \dot{F} \dot{\psi} + 6 \dot{F}H\theta +
 \dot{\phi}\dot{\delta{\phi}} 
\nonumber \\
&\;&\;\;\;-\ddot{\phi}\delta\phi - 3H\dot{\phi}\delta\phi
 -3H\dot{\delta F}+3\dot{H}\delta F +3H^2\delta F -\frac{k^2}{a^2}
 \delta F = 0 
\label{eq:pert00}\\
& & 6F\ddot{\psi}+12F\dot{H}\theta + 6 FH\dot{\theta} + 12 FH\dot{\psi} + 12 FH^2\theta - 2F\frac{k^2}{a^2}\theta \nonumber\\
&\;&\;\;\;+3\dot{F}\dot{\psi} + 6\dot{F}H\theta + \dot{F}\dot{\theta} + 4 \dot{\phi}^2\theta + 6 \theta \ddot{F}\nonumber\\
&\;&\;\;\; -4\dot{\phi}\dot{\delta\phi}-2\ddot{\phi}\delta\phi - 6H\dot{\phi}\delta\phi - 3\ddot{\delta F}- 3H\dot{\delta F}+6H^2\delta F-\frac{k^2}{a^2}\delta F = 0 \label{eq:pertii}\\
& & \ddot{\delta \phi} + 3H\dot{\delta \phi} - \frac{1}{2}f_{\phi\phi} + V_{\phi \phi}\delta \phi + \frac{k^2}{a^2}\delta \phi - 3 \dot{\phi}\dot{\psi} -6H\dot{\phi}\theta -\dot{\phi}\dot{\theta}  \nonumber \\ 
&\;&\;\;\; - 2\ddot{\phi}\theta + 3 F_\phi \ddot{\psi} + 6F_{\phi} \dot{H} \theta + 3 H F_{\phi} \dot{\theta} + 12 F_{\phi}H\dot{\psi} +12 F_{\phi}H^2 \theta + \nonumber\\
& & 2 F_{\phi} \frac{k^2}{a^2}\psi - F_{\phi} \frac{k^2}{a^2}\theta = 0
\label{eq:perteomphi}\\
& & \delta F - F_{\phi}\delta \phi + F_R\delta R = 0 \label{eq:pertF1}
\end{eqnarray}
\label{eq:pertscalar} 
\end{subequations}
where
\[
\delta R = -6\ddot{\psi}-12 \dot{H} \theta - 6 H \dot{\theta} - 24 H \dot{\psi} - 24 H^2 \theta - 4\frac{k^2}{a^2}\psi + 2 \frac{k^2}{a^2} \theta 
\]
It is important to note that only three of the above six equations are independent 
and we have rewritten the perturbation of $F$ in Eq. (\ref{eq:pertF1}) as an independent equation. 

The tensor perturbations in the Fourier space are given by: 
\begin{equation}
\ddot{C}^\alpha_\beta + \left(\frac{\dot{F}}{F} + 3H\right)\dot{C}^\alpha_\beta+\frac{k^2}{a^2}C^\alpha_\beta=0
\label{eq:perttensor}
\end{equation}
In the rest of this section, we obtain the scalar and tensor power spectrum for the exact de Sitter model. The reason is two fold: First, for an inflationary model with an exit, the scale factor evolution can only be obtained numerically and hence, the power-spectrum can not be evaluated exactly. Second, the perturbations equations are complicated and for the exact de sitter, the perturbation equations can be simplified and we can obtain analytical solutions in the super-Hubble scales.  

We obtain the scalar power-spectrum in the limit $p \ll 1$, however, the tensor power-spectrum is obtained for any $p$ that leads to inflation. 

\subsection{Scalar Power Spectrum}

Since the analysis is in Jordan frame, the quantity we need to evaluate inorder to  compare with the observations is 3-Curvature perturbation (${\cal R}$) which is given by: 
\begin{equation}
\label{eq:3Curvaturedef}
{\cal R} = \psi + \frac{H}{\dot{\phi}} \delta\phi
\end{equation}
Before we proceed with the evaluation, we would like to point that  ${\cal R}$ is conserved at large scales. The constancy of ${\cal R}$ is a consequence of the local energy conservation and is valid for any relativistic theory of gravity~\cite{wands2000new,lyth2005general}. 
In the rest of the analysis, we will not be including the entropy perturbations as these will vanish at the super-Hubble scales. 

As the equations (\ref{eq:pertscalar}) are highly coupled, non-linear and higher-order, we need to follow different strategy to obtain the differential equation for the 3-curvature perturbation. 

First, is to obtain a solution to the differential equation of the combined variable 
$\theta + \psi \equiv \Theta $.  Physically, $\Theta$ is the Bardeen potential 
in the Einstein frame. From Eqs.(\ref{eq:pertinotj},\ref{eq:pert0i},\ref{eq:pert00}), the differential equation for $\Theta$ is given by: 
\begin{equation}
\begin{aligned}
& F\ddot{\Theta} + \left( 3\dot{F}  + H \, F - 
-\frac{2F\ddot{\phi}}{\dot{\phi}} \right)  \dot{\Theta} 
+ \left(  \frac{k^2}{a^2} F  - \ddot{F} -\frac{2FH\ddot{\phi}}{\dot{\phi}} 
+ \frac{2 \dot{F} \ddot{\phi}}{\dot{\phi}} + H\dot{F} + 4\dot{H} F
\right) \Theta \\
&= \left( \dot{\phi}^2  + 6 F \dot{H} - 3\dot{F}H  - 3 \ddot{F}
+ \frac{6 \dot{F} \ddot{\phi}}{\dot{\phi}} \right) \theta 
\end{aligned}
\label{eq:master1}
\end{equation}
For the exact analytical solution in the previous section ($H$ is a constant), Eq.~(\ref{eq:master1}) becomes:
\begin{equation}
\label{eq:master1dS}
\ddot{\Theta} + H_D(1- 4 p) \dot{\Theta} + \frac{k^2}{a^2}\Theta - 4 \, p \, H_D^2 \, (1-p)\theta = 0
\end{equation}
In the small wavelength limit $\frac{k}{a}\gg1$,  the last two terms in the LHS can be approximated as 
\[
\left( \frac{k^2}{a^2}  - 4 \, p \, H_D^2 \, (1-p) \right) \theta + 
\frac{k^2}{a^2} \psi \simeq  \frac{k^2}{a^2} \Theta \, .
\]
we then have, 
\begin{equation}
\label{eq:master1dSapprox}
\ddot{\Theta} + H_D(1- 4 p) \dot{\Theta} + \frac{k^2}{a^2}\Theta \simeq 0
\end{equation}
Second, rewrite $\delta \phi$ in terms of $\Theta$. Using Eqs. (\ref{eq:pertinotj},\ref{eq:pert0i}), for $p \ll 1$, we have 
 \begin{equation}
 \delta\phi=\frac{\phi_0}{2 H_D} \, e^{-p \, H_D \, t} 
 \left( \dot{\Theta}+H_D\Theta \right)
 \label{eq:phi1}
 \end{equation}
Third, using the combination of Eqs. (\ref{eq:pert0i}, \ref{eq:pertinotj}) and (\ref{eq:pertF1}), we obtain $\psi$, $\theta$ and $\delta F$ in terms of $\Theta$, i. e.,

\begin{subequations}
\label{eq:var-Theta}
\begin{eqnarray}
\theta&=&\frac{2}{3}\frac{1}{\frac{k^2}{a^2}}\ddot{\Theta}+\frac{1}{\frac{k^2}{a^2}} \dot{\Theta}\left(H_D -\frac{1}{12\alpha H_D}\right) + \frac{2}{3}\Theta \\
\psi &=& \frac{1}{3}\Theta - \frac{2}{3}\frac{1}{\frac{k^2}{a^2}}\ddot{\Theta}-\frac{1}{\frac{k^2}{a^2}} \dot{\Theta}\left(H_D -\frac{1}{12\alpha H_D}\right) \\
\delta F &=& n \phi_0^2 e^{-2pH_Dt}\left(\frac{1}{6}\Theta+\frac{2}{3}\frac{1}{\frac{k^2}{a^2}}\ddot{\Theta}+\frac{1}{\frac{k^2}{a^2}} \dot{\Theta}\left(H_D -\frac{1}{12\alpha H_D}\right) \right)
\end{eqnarray}
\end{subequations}
Fourth, solve the differential equation (\ref{eq:master1dSapprox}) in the small wavelength limit to obtain $\Theta$, i.e.,
\begin{equation}
\Theta=e^{(4 p -1)H_Dt/2} U_1
\end{equation} 
where 
\[
U_1=C_1H_{\frac{1}{2}-2
 p}^{(1)}\left(\frac{ke^{-H_Dt}}{a_0H_D}\right)+C_2H_{\frac{1}{2}-2
 p}^{(2)}\left(\frac{ke^{-H_Dt}}{a_0H_D}\right)
 \]
Fifth, we need to reduce the order of the differential equation of ${\cal R}$. From Eqs. (\ref{eq:var-Theta}), it is clear that the ${\cal R}$ contains higher derivatives of $\Theta$. Interestingly, it can be shown that, $\ddot{\Theta}$ is linear in $\Theta$ and $\dot{\Theta}$, see Eq.~\ref{eq:master1dSapprox}. After a long calculation and setting the Bunch-Davies vacuum, in the limit  $\frac{k}{a}\gg1$ the Jordan frame curvature perturbation $\mathcal{R}_<$ is given by:
\begin{equation}
\mathcal{R}_< \,=\,\frac{H_D}{2a\sqrt{k}}e^{-ik\eta}
\end{equation}
In the large scale limit, i.e. $\frac{k}{a}\rightarrow 0$, we can see that $\mathcal{R} = constant$ is a solution. Hence, we have 
\begin{equation}
\mathcal{R_>}= {C} \, .
\end{equation}
Matching the small wavelength and large wavelength solutions at $|k\eta|=2\pi$ we have ${C} =\frac{\sqrt{2} H_D \pi}{k^{3/2}}$. 
For $p \ll 1$,  the scalar power-spectrum is constant and is given by:
\begin{equation}
\mathcal{P}_\mathcal{R}={H_D^2} \, .
\end{equation}

\subsection{Tensor Power Spectrum}

Following Ref.~\cite{hwang2000conserved}, we can obtain the following equation of motion for the tensor perturbation for exact de Sitter solution:
\begin{equation}
\ddot{C}^\alpha_\beta + \left(-2pH_D + 3H_D\right)\dot{C}^\alpha_\beta+\frac{k^2}{a^2}C^\alpha_\beta=0
\label{eq:perttensor_par}
\end{equation} 
Defining  $C^{\alpha}_\beta=\nu_g/z_g$ and $z_g= a\phi_0 \sqrt{1+24 \alpha H_D^2} e^{-pH_Dt}$, we 
have 
\begin{equation}
\nu_g'' + \left(k^2-\frac{z_g''}{z_g}\right)\nu_g=0
\end{equation}
The solution to the above differential equation is again a sum of Hankel function:
\begin{equation}
\nu_g=\sqrt{-\eta}(\tilde{C}_1H_{3/2-p}^{(1)}(-k\eta) +\tilde{C}_2 H_{3/2-p}^{(2)}(-k\eta))
\end{equation}
Setting the initial state to be Bunch-Davies vacuum, we have $\tilde{C}_2=0$ and $\tilde{C}_1=\sqrt{\frac{\pi}{4}}$. Hence for tensor perturbation $C^{\alpha}_{\beta}$, we have:

\begin{equation}
\nu_g= \sqrt{\frac{\pi}{4}}\sqrt{-\eta}H_{3/2-p}^{(1)}(-k\eta)
\end{equation}
The Tensor power spectrum is given by $\mathcal{P}_{{g}}= 8\frac{k^3}{2\pi^2}|\mathcal{C^{\alpha}_{\beta}}|^2$ and is:
\begin{equation}
\mathcal{P}_{g}=8\left(\frac{k}{k_*}\right)^{2p} \frac{2^{-2p}}{4\pi^2}H_D^{2} \left(\frac{\Gamma(3/2-p)}{\Gamma(3/2)}\right)^2e^{2 p H_Dt_*} {\phi_0}^2 \left({1+24 \alpha H_D^2}\right)
\end{equation}
Tensor spectral index, $n_T=2p$, which means that for a decaying scalar field the spectrum obtained is blue tilted.

\section{Discussions}
\label{sec:Conclusions}
In this work, in Jordan frame, we have obtained an exact inflationary model for an $f(R,\phi)$ model. The scalar field is massive, non-minimally coupled to $f(R)$ and does not have self-interacting potential. The scalar field potential is consistent with standard model of particle physics. We have shown analytically and numerically that the model has an inflationary solution with an exit. For large number of e-foldings, the inflationary model behaves close to de Sitter. 

It is important to note that the scalar perturbations cannot be evaluated analytically for any value of $p$. To verify the validity of the model with the CMB observations, we have obtained the scalar power spectrum for $p \ll1$. Under this limit, the scalar power spectrum is nearly scale-invariant. We need to numerically evaluate for general $p$ to be constraint the model with the current PLANCK data.

For a more precise calculation which was possible in the case of tensor perturbation, we have shown that the obtained spectrum have a blue tilt -- for the exact constant $H$ solution. The requirement that the blue tilt must be very small \cite{ade2014planck2} constrains the parameter $p \ll 1$. The running of spectral index is expected to be negative for our model. In order to show that we need to obtain the power-spectrum numerically, which is currently under investigation. 

\section{Acknowledgements}
JM is supported by UGC Senior Research Fellowship, India and JPJ is supported by CSIR Junior Research Fellowship, India. The work is supported by Max Planck-India Partner Group on Gravity and Cosmology.  

\appendix 
\section{Details of the Analytical approach for the saddle point}
\label{app:analytical}
Jacobi matrix, $J_{i j}$, is given by:
\begin{equation}
\arraycolsep=1.4pt\def\arraystretch{1.5}
\setlength\arraycolsep{10pt}
J_{ij}=
\left[ \begin {array}{ccc}
 0&1&0\\ \noalign{\medskip}1/6\,{\frac {72
\,p\alpha\,{H_D}^{2}+p-1}{\alpha}}&2\,pH_D-3\,H_D&1/9\,{\frac
 {-1-24\,\alpha
\,{H_D}^{2}+p+24\,p\alpha\,{H_D}^{2}}{\alpha}}\\ \noalign{\medskip}3\,{
\frac { \left( -3+2\,p \right) pH_D}{1+2\,p}}&-3\,{\frac
 {p}{1+2\,p}}&2
\,pH_D-3\,H_D\end {array} \right] 
\label{eq:JacobiMatrix}
\end{equation} 
Let the eigen value and eigen vector of $J$ be $\lambda_i$ and $u_i$.
Then the phase space trajectory is given by:
\begin{equation}
\delta{v_i}=\sum_{i=1}^{i=3}c_i u_i e^{(\lambda_i t)}
\end{equation}
\vspace{2 mm}
where
\small
\[
\lambda_i=\left[ \begin{array}{c}
\lambda_1 \\ \lambda_2 \\ \lambda_3
\end{array}\right]=
\left[ \begin {array}{c} 2\,pH_D-3\,H_D\\ \noalign{\medskip}\,{\frac {
-9\,H_D\alpha-12\,H_D\alpha\,p+12\,H_D\alpha\,{p}^{2}+\sqrt
 {81\,{H_D}^{2}{
\alpha}^{2}+936\,{H_D}^{2}{\alpha}^{2}p+1944\,{H_D}^{2}{\alpha}^{2}{p}^{2}
+864\,{H_D}^{2}{\alpha}^{2}{p}^{3}+144\,{H_D}^{2}{\alpha}^{2}{p}^{4}-6\,p
\alpha-6\,\alpha+12\,{p}^{2}\alpha}}{ 6\left( 1+2\,p \right)
 \alpha}}
\\ \noalign{\medskip}-\,{\frac
 {9\,H_D\alpha+12\,H_D\alpha\,p-12\,H_D
\alpha\,{p}^{2}+\sqrt
 {81\,{H_D}^{2}{\alpha}^{2}+936\,{H_D}^{2}{\alpha}^{2
}p+1944\,{H_D}^{2}{\alpha}^{2}{p}^{2}+864\,{H_D}^{2}{\alpha}^{2}{p}^{3}+
144\,{H_D}^{2}{\alpha}^{2}{p}^{4}-6\,p\alpha-6\,\alpha+12\,{p}^{2}\alpha
}}{ 6\left( 1+2\,p \right) \alpha}}\end {array} \right] \]
\vspace{2 mm}
and
\normalsize
\[\arraycolsep=1.4pt\def\arraystretch{2.2}
u_1=\left[ \begin{array}{c}
-\frac{2}{3}\frac{-1-24 \alpha H_D^2 + 24
 \alpha p H_D^2+p}{72 p H_D^2 \alpha + p-1}
 \\ -\frac{2}{3}\frac{-1-24 \alpha H_D^2 + 24 \alpha p H_D^2+p}{72
 p H_D^2 \alpha + p-1} \lambda_1 \\ 1
\end{array}\right];                  
\quad u_2 =\left[ \begin{array}{c}
 \frac{1}{18}\frac{1-p-120 \alpha p
 H_D^2 - 96 \alpha p^2
 H_D^2}{p\alpha\lambda_2(\lambda_2-\lambda_1)}
 \\ \frac{1}{18}\frac{1-p-120 \alpha p H_D^2 - 96 \alpha p^2
 H_D^2}{p\alpha\lambda_2(\lambda_2-\lambda_1)} \lambda_2 \\ 1
\end{array}\right]; 
\quad u_3 =\left[ \begin{array}{c}
\frac{1}{18}\frac{1-p-120 \alpha p
 H_D^2 - 96 \alpha p^2
 H_D^2}{p\alpha\lambda_3(\lambda_3-\lambda_1)}
 \\ 
\frac{1}{18}\frac{1-p-120 \alpha p H_D^2 - 96 \alpha p^2
 H_D^2}{p\lambda_3(\lambda_3-\lambda_1)} \lambda_3 \\
1 
\end{array}\right]
\]
$c_i$'s are constants whose values has to be fixed from initial values of Hubble parameter ($H_i$) and the initial value of $\dot{\phi}/\phi$ ($\Delta_i$).\\ \\ \\

\bibliography{bibdata.bib}

\end{document}